\documentclass[nofootinbib,aps,twocolumn,letterpaper,preprintnumbers,superscriptaddress]{revtex4-2}
\usepackage{amsmath}
\usepackage{eurosym}
\usepackage{amssymb}
\usepackage{graphicx,epsfig}
\usepackage{color}
\usepackage{braket}
\usepackage{booktabs}
\usepackage{dcolumn}
\usepackage{multirow}
\usepackage{amsmath}
\usepackage{amssymb}
\usepackage{subfigure}
\usepackage{amsfonts}
\usepackage[left=1.5cm,right=1.5cm,top=1.5cm,bottom=1.5cm]{geometry}
\usepackage[usenames,dvipsnames,svgnames]{xcolor}  
\usepackage{hyperref}
\definecolor{darkgreen}{rgb}{0.1,0.6,0.1}
\definecolor{darkblue}{rgb}{0,0,0.3}
\definecolor{darkred}{rgb}{0.7,0,0}

\definecolor{light gray}{RGB}{220,220,220}
\definecolor{dark purple}{RGB}{108,0,217}
\definecolor{pink}{RGB}{190,20,100}
\definecolor{orang}{RGB}{193,63,0}
\definecolor{green}{RGB}{11,98,17}
\definecolor{darkpink}{RGB}{153,0,76}
\definecolor{bluegreen}{RGB}{0,102,102}
\definecolor{greenlagan}{RGB}{0,102,0}
\definecolor{redgreen}{RGB}{102,102,0}
\definecolor{Redgreen}{RGB}{153,76,0}
\definecolor{vividviolet}{rgb}{0.62, 0.0, 1.0}
\definecolor{amaranth}{rgb}{0.9, 0.17, 0.31}
\definecolor{palatinateblue}{rgb}{0.15, 0.23, 0.89}
\definecolor{brightpink}{rgb}{1.0, 0.0, 0.5}
\definecolor{cornflowerblue}{rgb}{0.39, 0.58, 0.93}
\definecolor{deepcarminepink}{rgb}{0.94, 0.19, 0.22}
\definecolor{radicalred}{rgb}{1.0, 0.21, 0.37}

\definecolor{beamer@PRD}{RGB}{46,48,146}
\hypersetup{
     breaklinks=true,
    pdfstartview={FitH},    
    colorlinks=true,       
    linkcolor=vividviolet,          
    citecolor=blue,        
    filecolor=magenta,      
    urlcolor=brightpink,           
    anchorcolor=cyan,      
    linktocpage=true
}

\begin{document}
{\vskip .1cm}
\date{\today}
\newcommand\be{\begin{equation}}
\newcommand\ee{\end{equation}}
\newcommand\bea{\begin{eqnarray}}
\newcommand\eea{\end{eqnarray}}
\newcommand\bseq{\begin{subequations}} 
\newcommand\eseq{\end{subequations}}
\newcommand\bcas{\begin{cases}}
\newcommand\ecas{\end{cases}}
\newcommand{\p}{\partial}
\newcommand{\f}{\frac}

\title{Evolution of Spherical Overdensities in Energy-Momentum-Squared Gravity}

\author{\bf Bita Farsi}
\email{Bita.Farsi@shirazu.ac.ir}
\affiliation{Department of
Physics, College of Sciences, Shiraz University, Shiraz 71454,
Iran}

\author{\bf Ahmad Sheykhi}
\email{asheykhi@shirazu.ac.ir}
\affiliation{Department of Physics, College of Sciences, Shiraz University, Shiraz 71454, Iran}
\affiliation{Biruni Observatory, College of Sciences, Shiraz University, Shiraz 71454, Iran}

\author {\bf{Mohsen Khodadi}}
\email{m.khodadi@hafez.shirazu.ac.ir}
\affiliation{Department of Physics, College of Sciences, Shiraz University, Shiraz 71454,
   Iran}
\affiliation{Biruni Observatory, College of Sciences, Shiraz University, Shiraz 71454, Iran}

\begin{abstract}
Employing the spherical collapse (SC) formalism, we investigate
the linear evolution of the matter overdensity for energy-momentum-squared gravity (EMSG), which in practical phenomenological terms, one may imagine as an extension of the $\Lambda$CDM model of cosmology. The underlying model, while still having a cosmological constant, is a non-linear material extension of the general theory of relativity (GTR) and includes correction terms that are dominant in the high-energy regime, the early universe. Considering the Friedman–Robertson–Walker (FRW) background in
the presence of a cosmological constant, we find the effects of
the modifications arising from EMSG on the growth of perturbations
at the early stages of the universe. Considering both possible negative and positive values of the model parameter of EMSG, we discuss its role in the evolution of the matter density contrast and growth function in the level of linear perturbations.
While EMSG leaves imprints distinguishable from $\Lambda$CDM, we
find that the negative range of the ESMG model parameter is not
well-behaved, indicating an anomaly in the parameter space of the
model. In this regard, for the evaluation of the galaxy cluster
number count in the framework of EMSG, we equivalently provide an
analysis of the number count of the gravitationally collapsed
objects (or the dark matter halos). We show that the galaxy
cluster number count decreases compared to the $\Lambda$CDM model.
In agreement with the hierarchical model of structure formation,
in EMSG cosmology the more massive structures are less
abundant, meaning that form at later times.
 \end{abstract}

\maketitle

\section{Introduction \label{Intro}}
It is not an exaggeration to say that the discovery of the current
cosmic acceleration of the universe is the most exciting
cosmological achievement in many decades. Generally speaking, it
implies that the universe is dominated by an enigmatic repulsive
force so-called dark energy (DE) with unusual physical properties,
or that the general theory of relativity (GTR) as the basis of
standard cosmology ($\Lambda$CDM) fails on cosmological scales
\cite{Wein}. This enigmatic force has inspired many researchers to reveal its unknown properties. In this regard, serving alternative theories of gravity that indeed are extensions of GTR are regarded as the most optimistic proposals to disclose the nature of the accelerating expansion of the universe. In addition to DE, there are other unresolved issues, such as the explanation of the missing matter is known as dark matter (DM) at the galactic and cosmological scales, and the existence of singularities at
high-energy regimes which keep open the way to going beyond GTR
(see \cite{Capo} for more knowing). In the review paper \cite{Nojiri}, one can find a systematic discussion on some standard issues and the latest developments of modified gravity in cosmology.

Usually, the theories beyond GTR build due by adding scalar
invariants (also their corresponding functions) in the geometric
section of the Einstein-Hilbert action. The most simple modification of GTR is the so-called $f(R)$ theory, in which the Ricci scalar $R$ takes different forms. The viable aspects of these alternative theories have been comprehensively discussed in the literature \cite {Cognola}.
In the continuation of this road, a generalization of $f(R)$ theory of gravity has been suggested in Ref. \cite{Harko} known as $f(R, T)$ gravity, where $T$ denotes the trace of the energy-momentum tensor (EMT). 
The outstanding
feature of $f(R, T)$ theories which are also known as minimal
curvature-matter coupling models \footnote{There is also the
non-minimal version of the interaction between geometry and matter
which is labeled as $f(R, T, R_{\mu \nu} T^{\mu \nu})$
\cite{Haghani:2013oma}.}, is that can provide worthy descriptions
for the late time cosmic acceleration and the interconnection of
DE, and DM, as well (see e.g.,
\cite{Myrzakulov,Harko2,Zaregonbadi}). Note that some of the classes of $f(R, T)$ gravity also suffer from inviability issues and can not provide a realistic cosmology \cite{Velten:2017hhf}.
Authors in \cite{Kat}, for the first time, developed theories exploiting the trace of EMT in the general form $f(R, (T_{\mu \nu}T^{\mu \nu})^n$
($0<n\leq1$), as the models violating the local/covariant energy-momentum conservation with the norm of EMT $T_{\mu\nu}T^{\mu \nu}$. Generally, these models are categorized under the name of energy-momentum-powered gravity (EMPG), so that in Ref. \cite{Kat} by focusing only on two cases $n=1/4$, and $1/2$ was showed that the addition of the norm of EMT to the action results in the Cardassian expansion \footnote{It is a kind of expansion for the flat universe consisting of only matter and radiation (without vacuum contribution) and still consistent with observations \cite{Freese:2002}.}. 
Further studies explicitly revealed to us that case $n<1/2$ will affect the dynamic of the universe in late time \cite{Akarsu:2017ohj} while $n>1/2$ is efficient to high-energy regimes \cite{Roshan, Barrow}. 
Case $n=1/2$ is very interesting in the sense that it affects the field equations independent of the energy density scale ( see the cosmological and gravitational implications of this case in \cite{Akarsu:2018aro, Akarsu:2023,Akarsu:2023e}).

With these prerequisites, since the topic under investigation in
this paper i.e., structure formation is influenced by the dynamic
governing the early universe, then we take the case $n=1$, which
the authors in Refs. \cite{Roshan, Barrow} have presented it as a
new covariant generalization of GTR relevant to the high energy
phase of the universe. The label of this model is energy-momentum-squared gravity (EMSG) so that inducing the quadratic contributions to gravity from the matter side makes the appearance of new forms of fluid stresses such as the scalar field and so on, unnecessary \cite{Akarsu}. 
In general, $f(R, (T_{\mu\nu}T^{\mu \nu})^n)$ gravities enjoy this feature that their modifications do not come from gravitational Lagrangian but from the matter Lagrangian, just unlike $f(R)$ theories.
In other words, the self-coupling of the matter this time instead of geometry is supposed to give us interesting cosmological outputs, in particular, concerning the early universe (case $n=1$). 
Studies
conducted in Refs. \cite{Roshan} and \cite{khodadi} (see also
relevant papers \cite{Sharif:2021,Sharif2023}) respectively
indicate that EMSG in the context of bounce \footnote{There is
another analysis also that shows EMSG can not replace the standard
initial singularity with a regular bounce, meaning that it still
suffers from geodesically incompleteness issue \cite{Barbar}.
However, this problem can be solved in case of the existence of a
vacuum energy density in EMSG \cite{Nazari:2020gnu}.} and emergent
scenarios is an alternative gravity without initial singularity.
It may be justified by the fact that EMSG, just similar to most quantum gravity approaches predicts a minimum length and a finite maximum energy density, leading to a circumvention of the initial singularity problem in GTR. 
Due to some modifications induced by
EMSG for the physics of the early universe, one can be considered
it a phenomenological effort to improve the usual paradigm in
$\Lambda$CDM-based cosmology. This idea, for several reasons, is justifiable. First, $\Lambda$CDM, due to the theoretical inconsistencies related to the cosmological constant, can not provide us with a self-consistent description of cosmic acceleration,
although it is successful in fitting a range of the observational data \cite{Weinberg:1988cp, Padmanabhan:2002ji}. 
Second, some data
analyses indicate incompatibilities between the $\Lambda$CDM
prediction and the constraints obtained from some local
observations, e.g., see
\cite{Zhao:2017cud,Bullock:2017xww,DiValentino:2017gzb}. Third, the Hubble tension that has recently been at the center of the
the attention of the cosmology community warns us of the possibility
of the invalidity of the $\Lambda$CDM picture of the early
universe \cite{Abdalla:2022yfr,Khodadi:2023ezj,Hu}. 
It would be helpful to note that despite the possibility of solving the cosmological constant issue at the late time via presenting quintessence DE, some studies show that it may make Hubble tension worse \cite{Lee:2022cyh, Banerjee:2020xcn}. 
With this background, one may potentially take the EMSG as phenomenological
extensions of the $\Lambda$CDM for the description of the early
universe, such that, the additional free parameter appeared in
Einstein’s equations (arising from the nonlinear matter
Lagrangian) finally can be constrained via cosmological
observations. About the late-time behavior of EMSG cosmology, in a
recent study \cite{HR} by taking into consideration the homogenous
and isotropic spacetime in the presence of the cosmological
constant for this theory, succeeded in deriving different
plausible scenarios of dark energy. Notably, one of these scenarios via presenting quintessence DE able to solve the cosmological constant issue at the late time.  In light of some
cosmological observation measurements \cite{Ranjit,Nozari} and
gravitational setups \cite{Akarsu,Nazari:2022xhv,Nazari} derived
different constraints for the free parameter of EMSG. 
Apart from these cases, we see EMSG in recent years subjected to evaluation in different contexts, see Refs.
\cite{Moraes,Faria,Bahamonde,Shahidi:2021,Kolonia,Khodadi:2022,Sharif:2022jiz,Sharif:2022aei,Sharif:2023uac}.

The structure formation of the universe is highly sensitive to the
accelerated expansion history of the universe since any change
affects the rate of formation and growth of collapsed structures
\cite{Abramo}. This is important because all galaxies, quasars,
and supernovae, in essence, come from collapsed structures, and
their distribution in size, space, and time is subsequently
affected. Indeed, large-scale structures in our current universe
are nothing but growing small fluctuations in the early universe.
This statement is valid for any source of changes in the expansion
history of the universe.  With this idea in mind, we are going to
study linear structure formation in cosmologies beyond standard
$\Lambda$CDM, in particular, EMSG cosmology.  The higher-order matter source terms appearing in the EMSG dynamical equations are expected to leave significant effects on the initial small fluctuations as well as the evolution of the structure. 
An appropriate approach to describe this evolution is known as Top-Hat SC formalism which addresses the growth of perturbations and subsequently structure formation \cite{Gunn:1972}. 
In this approach, one considers a homogeneous and spherical symmetric
perturbation in an expanding background and describes the growth of perturbations in a spherical region using the same Friedmann equations for the underlying theory of gravity
\cite{Planelles,Ziaie,Ziaie2,Farsi,Farsi2,Farsi3}. More precisely,
the Top Hat describes a homogeneous overdensity sphere that is, in
principle, modeled by a discrete Friedman-Robertson-Walker (FRW)
closed universe lying in an external FRW  universe, with flat
spatial curvature \cite{Munoz:2023rwh}. The radius of this overdensity sphere expands at a rate slower than the background, gradually slowing down so that it finally reaches its possible
the maximum size, and then reversely shrinks into itself to collapse.
Notably, our work here differs from \cite{Alvarenga2}, which has
utilized the Newtonian gauge since we instead serve SC formalism
to survey the evolution of perturbations. It is worth noting that recently in the framework of EMSG, by investigating the dynamics of SC for a spherically symmetric configuration such as a star has been addressed the stability of self-gravitating objects \cite{Sharif:2021uyc}.

The outline of this paper is as follows. In Sec. \ref{FEC}, we
provide a review on EMSG and derive the corresponding Friedmann
equations. In Sec. \ref{SEMS}, using the spherically collapse
approach, we explore the growth of matter perturbation in the
background of flat EMSG-based cosmology. In Sec. \ref{MN}, for the
existing cosmology framework we address the mass function and
number count of the collapsed objects. We also devote the
conclusion and discussion to the last Sec. \ref{CD}. In what
follows we work in the units $\hslash=c=\kappa=1$.
\section{Modified dynamic equations of EMSG}  \label{FEC}
In Ref. \cite{Barrow}, the modified action of the EMSG model in the presence of the cosmological constant $\Lambda$, presented in the following form
\begin{eqnarray}\label{action}
    S_{EMSG}=\frac{1}{2 \kappa}\int{\sqrt{-g}\left\lbrace F(R,T^{2})-2\Lambda +L_{m}\right\rbrace d^{4}x},
\end{eqnarray}
with $F(R, T^{2})=R+\eta T^{2}$, where $\kappa=8\pi G/c^4$, and
$T^{2}\equiv T_{\mu \nu} T^{\mu \nu}$ is the scalar formed from
the square of EMT. The action of matter also labels with $L_{m}$.
Here, $\eta$ is the model parameter denoting the coupling between
matter and geometry with SI unite $ s^4kg^{-2}$. Without going to
details, the Einstein's equation reads as
\begin{eqnarray}
    &G_{\mu \nu}&+\Lambda g_{\mu \nu}=\kappa \left((\rho +p) u_{\mu} u_{\nu} +p g_{\mu \nu} \right)\nonumber\\
    && +\eta\left(\dfrac{1}{2} (\rho^{2} +3 p^{2})g_{\mu \nu}+(\rho+p)(\rho +3p) u_{\mu} u_{\nu}\right).
    \label{field2}
\end{eqnarray}
The first term in the right-hand side of the equation above shows that the matter content described by a perfect fluid with the standard EMT $T_{\mu\nu}$, containing the energy density $\rho$, and the pressure $p$ which are connected via equation-of-state parameter $\omega=p/\rho$.
As is clear from the second term, EMSG induces the quadratic contributions to gravity from matter terms which causes the standard EMT is no longer conserved locally i.e., $\nabla_{\mu}T^{\mu\nu}\neq0$ (in \cite{khodadi} one can find more discussion on handling this issue). Note that here for perfect fluid utilized the standard Lagrangian density $L_{m}=p$.

The modified Friedmann equations for the spatially homogeneous and flat FRW metric
\begin{equation}
    ds^2=-dt^2+a^2(t)\left[dr^2+r^2(d\theta^2 +sin^{2}\theta d\phi^{2})\right],\label{metric}
\end{equation}
take the following forms \cite{Barrow}
\begin{eqnarray}\label{GeneralFried1}
    \left( \dfrac{\dot{a}}{a}\right) ^{2}=\dfrac{\Lambda}{3}+\dfrac{\rho}{3}+\dfrac{\eta \rho^{2}}{6}\left(3\omega^2+8\omega +1 \right),
\end{eqnarray}
\begin{eqnarray}\label{GeneralFried2}
    \dfrac{\ddot{a}}{a}=\dfrac{\Lambda}{3}-\dfrac{1+3\omega}{6}\rho -\dfrac{\eta\rho^2}{3}\left(3\omega^2 +2\omega+1\right)~.
\end{eqnarray}
An interesting point in the above modified Friedmann equations is
that in case of choosing $\eta<0$ ($\eta>0$), one can recover the
effective Friedmann equations released in loop quantum cosmology
\cite{Ashtekar:2011ni} (braneworld cosmologies
\cite{Brax:2003fv}). In this regards, by differentiating the
Friedmann equations, the relevant continuity equation acquires as
\begin{eqnarray}\label{Generalcontinuity}
    \dot{\rho}=-3 \dfrac{\dot{a}}{a}\rho (1+\omega)\dfrac{1+\eta \rho(1+3\omega)}{1+\eta \rho (3\omega^{2}+8 \omega +1)},
\end{eqnarray}
which $\rho$ indeed is the energy density of both baryonic and DM.

By taking the case of dust matter in the form of $p=p_{m}=0,~ \omega_{m}=0,~ \rho=\rho_{m}$, so Eqs. (\ref{GeneralFried1}), and (\ref{GeneralFried2}) can be reexpress
\begin{eqnarray}\label{Fried1}
    3H^{2}=\Lambda +\rho_{m}+\dfrac{\eta}{2}\rho_{m}^{2},
\end{eqnarray}
\begin{eqnarray}\label{Fried2}
    2\dot{H}+3H^{2}=\Lambda - \dfrac{\eta}{2}\rho_{m}^{2}.
\end{eqnarray}
We note that in GTR limit (i.e., $\eta=0 $), the standard
Friedmann equations will be recovered from Eqs. (\ref{Fried1}) and
(\ref{Fried2}). The continuity equation (\ref{Generalcontinuity})
also for the underlying case leads to the usual behavior for the
matter density as
\begin{equation}
    \dot{\rho_m}+3H\rho_{m}=0. \label{continuty}
\end{equation}
The energy density of the pressureless matter ($p_m=0$) can be
obtained as $\rho_{m}=\rho_{m,0}a^{-3}$. The density parameters
are given by
\begin{eqnarray}
    &&\Omega_{\Lambda}=\frac{\Lambda}{3H^{2}}\label{OmLambda},\\&&
    \Omega_{m}=\frac{\rho_{m}}{3H^{2}},\label{OMmatter}\\ &&
    \Omega_{\eta}=\dfrac{\eta \rho_{m}^{2}}{6H^{2}}.\label{Ometa}
\end{eqnarray}
By substituting Eqs. (\ref{OmLambda})--(\ref{Ometa}) in Eq. (\ref{Fried1}) we obtain the following expression for the normalized Hubble parameter
\begin{equation}
E(z)=\dfrac{H(z)}{H_{0}}=\sqrt{ \Omega_{\Lambda,0}+\Omega_{m,0}(1+z)^{3}+\Omega_{\eta ,0}(1+z)^{6}}.
    \label{FriedEZ}
\end{equation}
By taking this fact into account that at the present-time Universe  $E(z=0)=1$, from Eq. (\ref{FriedEZ}), one gets the following relation between density parameters
\begin{equation}
\Omega_{m,0}+\Omega_{\Lambda,0}=1-\Omega_{\eta,0} .
    \label{Omega0}
\end{equation}
By relaxing $\eta$, the standard density equation is recovered,
as expected. In light of the strong evidence that the Universe is Euclidean, and the total density parameter is $\Omega\equiv\Omega_{m}+\Omega_{\Lambda}=1$, thereby, the value of the EMSG model parameter $\eta$ should be tiny.

\begin{figure}[t]
\epsfxsize=7.5cm \centerline{\epsffile{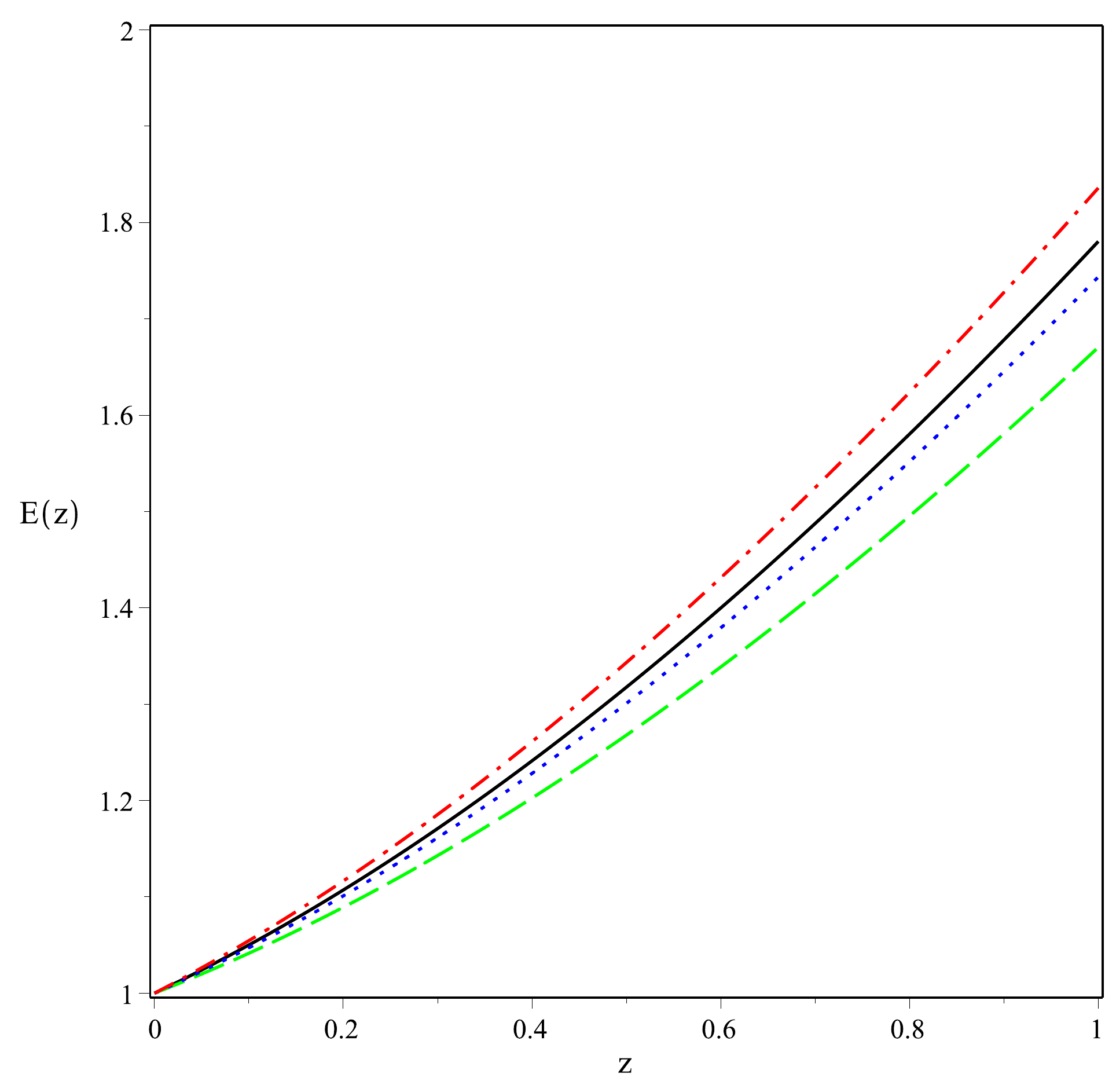}}
\caption{The behavior of the normalized Hubble rate $E(z)$ as a function of redshift $z$ for different values of EMSG model parameter $\eta$: $\eta=0$ ($\Lambda$CDM, black-solid), $\eta=10^{-5}$ (blue-dotted) $\eta=10^{-4}$ (green-dashed), and $\eta=-10^{-5}$ (red-dashed-dotted).} \label{Fig1}
\end{figure}

In Fig. \ref{Fig1} plotted $E(z)-z$ for different values of $\eta$.
As we can see, in EMSG cosmology, the normalized Hubble
parameter decreases by moving from negative values of the model parameter $\eta$ to positive ones. Also, for $\eta>0 ~(\eta < 0)$, the normalized Hubble parameter has a lower (steeper) slope, meaning that the rate of expansion of the universe becomes slower (faster) relative to the standard $\Lambda$CDM model.

From Eqs. (\ref{Fried1}), (\ref{continuty}), and (\ref{OMmatter}), the evolution of the density abundance $\Omega_{m}$, can be written
\begin{eqnarray}
    &&\Omega_{m}(z)\equiv \dfrac{\rho_{m}}{3H^{2}}= \dfrac{\rho_{m,0}(1+z)^3}{\rho_{m,0}(1+z)^3\left[1+\dfrac{\eta}{2} \rho_{m,0}(1+z)^3 \right]+\Lambda }\nonumber\\&&
    =\dfrac{\Omega_{m,0}}{\Omega_{m,0}+(1-\Omega_{m,0}-\Omega_{\Lambda,0})(1+z)^3+\Omega_{\Lambda,0}(1+z)^{-3}}.\nonumber\\
    \label{OMmatter2}
\end{eqnarray}

In a similar manner, from Eqs. (\ref{Fried1}), (\ref{continuty}), and (\ref{OmLambda}),
the evolution of the density abundance $\Omega_{\Lambda}$ takes the following form
\begin{eqnarray}
    &&\Omega_{\Lambda}(z)\equiv \dfrac{\Lambda}{3H^{2}}= \dfrac{\Lambda}{\rho_{m,0}(1+z)^3\left[1+\dfrac{\eta}{2} \rho_{m,0}(1+z)^3 \right]+\Lambda }\nonumber\\&&
    =\dfrac{\Omega_{\Lambda,0}(1+z)^{-3}}{\Omega_{m,0}+(1-\Omega_{m,0}-\Omega_{\Lambda,0})(1+z)^3+\Omega_{\Lambda,0}(1+z)^{-3}}.\nonumber\\
    \label{OMLambda2}
\end{eqnarray}
Also, using Eqs. (\ref{Fried1}), (\ref{continuty}), and (\ref{Ometa}), for the evolution of the density abundance $\Omega_{\eta}$ we have
\begin{eqnarray}
    &&\Omega_{\eta}(z)\equiv \dfrac{\eta\rho_{m}^2}{6H^{2}}= \dfrac{\eta}{2}\dfrac{\rho_{m,0}^2(1+z)^6}{\rho_{m,0}(1+z)^3\left[1+\dfrac{\eta}{2} \rho_{m,0}(1+z)^3 \right]+\Lambda }\nonumber\\&&
    =\dfrac{\Omega_{\eta,0}(1+z)^{6}}{\Omega_{m,0}(1+z)^{3}+\Omega_{\eta,0}(1+z)^{6}+\Omega_{\Lambda,0}}\nonumber\\&&
    =\dfrac{(1-\Omega_{m,0}-\Omega_{\Lambda,0})(1+z)^3}{\Omega_{m,0}+(1-\Omega_{m,0}-\Omega_{\Lambda,0})(1+z)^3+\Omega_{\Lambda,0}(1+z)^{-3}}.\nonumber\\
    \label{OMeta2}
\end{eqnarray}

\begin{figure}[t]
\epsfxsize=7.5cm \centerline{\epsffile{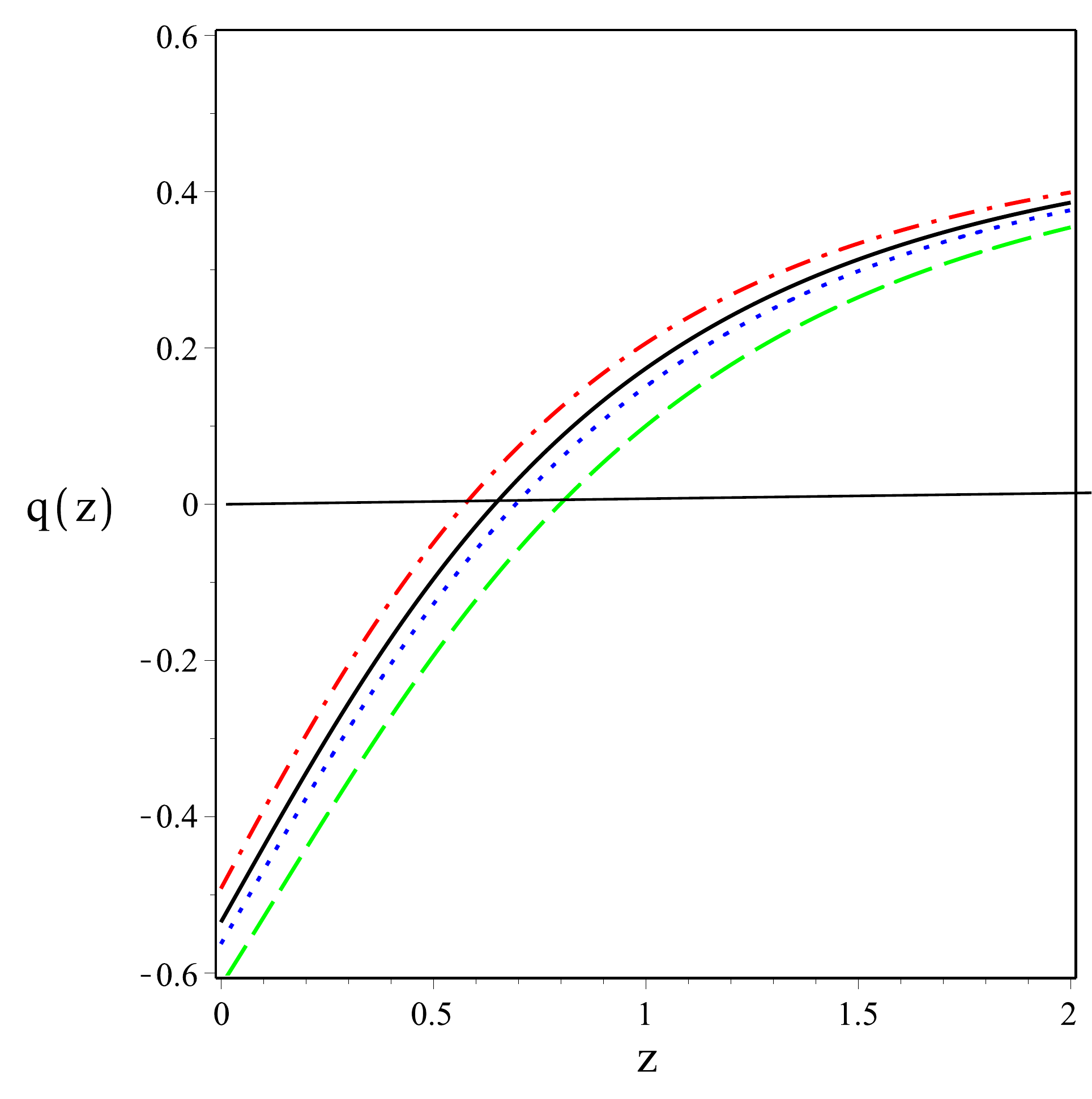}}
\caption{The behavior of the deceleration parameter $q(z)$ as a function of redshift $z$ for different values of the EMSG model parameter $\eta$: $\eta=0$ ($\Lambda$CDM, black-solid), $\eta=10^{-5}$ (blue-dotted) $\eta=10^{-4}$ (green-dashed), and $\eta=-10^{-5}$ (red-dashed-dotted).} \label{Fig2}
\end{figure}

The deceleration parameter in terms of the redshift can be written as
\begin{eqnarray}
    &&q=-1-\dfrac{\dot{H}}{H^2}=-1+\dfrac{(1+z)}{H(z)}\dfrac{dH(z)}{dz} \nonumber\\
    &&=\dfrac{-2\Omega_{\Lambda,0}+\Omega_{m,0}(1+z)^3+4(1-\Omega_{\Lambda,0}-\Omega_{m,0})(1+z)^6}{2\Omega_{\Lambda,0}+2\Omega_{m,0}(1+z)^3+2(1-\Omega_{\Lambda,0}-\Omega_{m,0})(1+z)^6}.\nonumber\\
    \label{deceleration}
\end{eqnarray}
One can see immediately that if $\eta=0, \Lambda=0$,
then Eq. (\ref{deceleration}), reproduces exactly GTR value i.e., $q=1/2$. We depict the behavior of the deceleration parameter $q$ for different $\eta$ in Fig. \ref{Fig2}. It reveals to us that the Universe experiences a transition from decelerating phase ($z > z_{tr}$) to the accelerating phase ($z < z_{tr}$), where $z_{tr}$ is the redshift at which the deceleration parameter vanishes.
We see that with decreasing $\eta$ (from positive to negative values), the phase transition between deceleration and acceleration takes place at lower redshifts. To clearly illustrate the role of the EMSG model parameter $\eta$,  the value of $z_{tr}$ are larger (smaller) than $\Lambda$CDM model for $\eta> 0~ (\eta < 0)$, meaning that for $\eta < 0$, this transition occurs later.
Overall, from the combination of the results of Figs. \ref{Fig1}, and \ref{Fig2}, it can be said that $\eta> 0~ (\eta < 0)$ causes the universe experiences an accelerated phase sooner (later) but with the expansion rate slower (faster) compared to $\Lambda$CDM model.

\section{Spherical Collapse (SC) in EMSG-setup}\label{SEMS}
\begin{figure}[t]
\epsfxsize=7.5cm \centerline{\epsffile{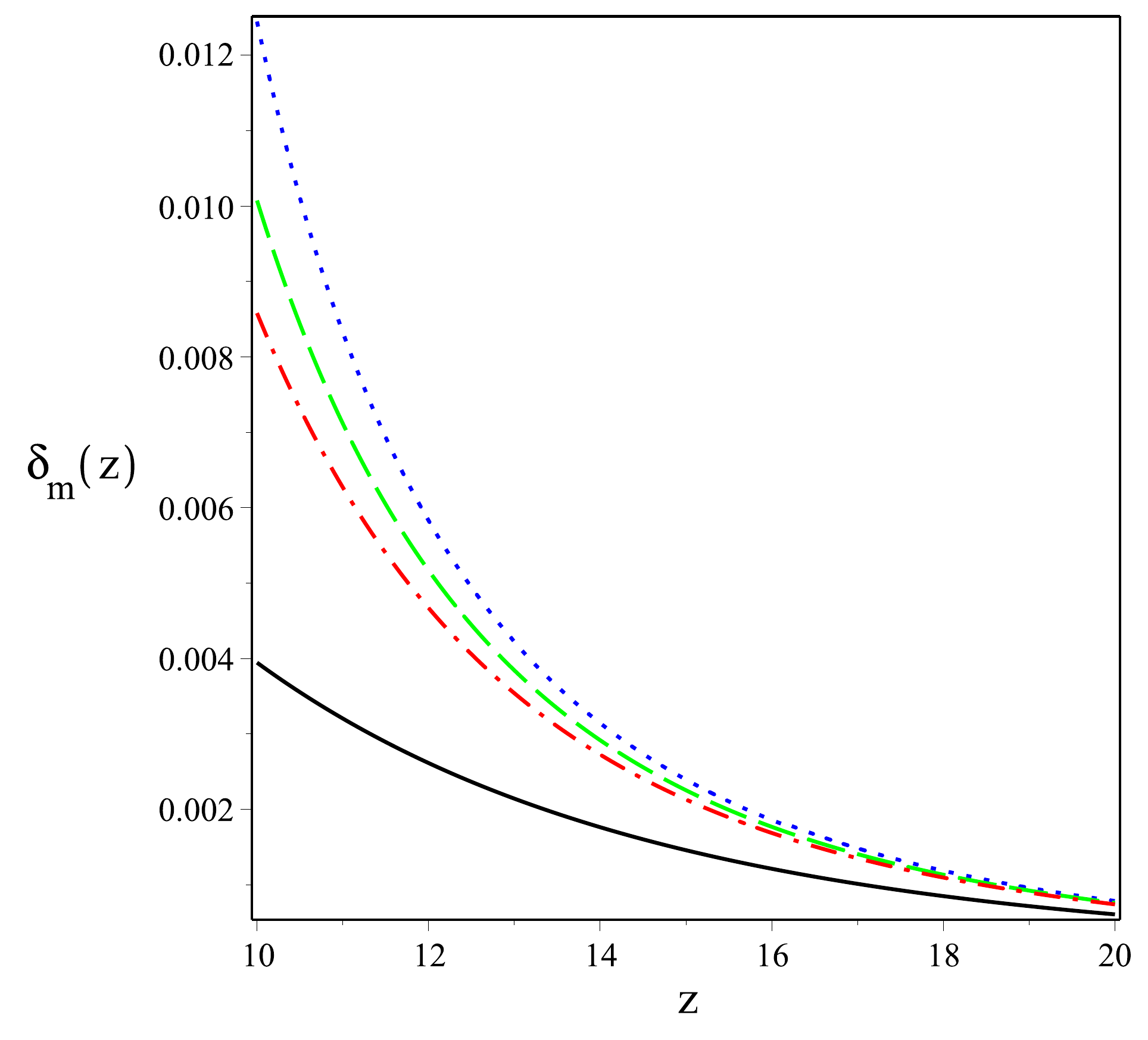}}
\caption{The evolution of the matter density contrast for different
values of $\eta$ during the evolution of the Universe. We have chosen
$\delta_{m}(z_{i})=0.0001$ , $z_{i} = 400$, with different values for the EMSG model parameter $\eta$: $\eta=0$ ($\Lambda$CDM, black-solid),  $\eta=10^{-6}$ (red-dashed-dotted), $\eta=10^{-4}$ (green-dashed), and $\eta=-10^{-6}$ (blue-dotted) .} \label{Fig3}
\end{figure}

By considering a dust fluid as the only constituent of the Universe
(i.e., $p = p_{m} = 0$), Eq. (\ref{continuty}) leads to
\begin{equation}
\dot{\rho}_{m}+3H\rho_{m}=0,
    \label{continuty1}
\end{equation}
with a solution in the form $\rho_{m}= \rho_{m,0}a^{-3}$ in which
$\rho_{m,0}$ denotes the energy density at the present time.
To investigate the growth of perturbations, we take a spherically symmetric perturbed cloud of radius $a_{p}$, with
a homogeneous energy density $\rho_{m}^{c}$. The SC model due to describing a spherical region with a top-hat profile and uniform density lets us write $\rho_{m}^{c}(t)=\rho_{m}(t)+\delta\rho_{m}$, at any time $t$ \cite{Ziaie2}. Here we are faceing with two cases $\delta\rho_{m}>0$ and $\delta\rho_{m}<0$. While for the former, this spherical region will eventually collapse under its gravitational force, for the latter, it will expand faster than the average Hubble expansion rate, subsequently generating a void.
In other words, these two cases happen in overdense, and underdense regions, respectively. Namely, in the matter-dominated universe, denser regions expand slower than the rest of the universe, which means if their density is large enough, they finally collapse and make clusters and other gravitational-bound systems \cite{Ryden}.
Like Eq. (\ref{continuty1}),  the continuity equation for the spherical region also takes the following form
\begin{equation}
    \dot{\rho}_{m}^{c}+3h\rho_{m}^{c}=0,~~~h=\dot{a}_{p}/a_{p} \label{continuty2}
\end{equation}
where $h$ refers to the local expansion rate of the spherical region with perturbed radius $a_p$. In line with our goal, let us start with definition of density contrast as \cite{Ryden}
\begin{equation}
    \delta_{m}=\dfrac{\rho_{m}^{c}}{\rho_{m}}-1=\dfrac{\delta\rho_{m}}{\rho_{m}},
    \label{delta}
\end{equation}
which $\rho_{m}^{c}$, $\rho_{m}$ are the energy density of spherical perturbed cloud and the background density, respectively. Note that the advantage of the above definition is that it allows us to work with dimensionless quantities. It, in essence, measures the deviation of the local fluid density from the background density.
Now, by taking the first and second time derivatives of Eq. (\ref{delta}), we respectively have
\begin{eqnarray}
    &&\dot{\delta}_{m}=3(1+\delta_{m})(H-h),\label{deltadot} \\
    &&\ddot{\delta}_{m}=3(\dot{H}-\dot{h})(1+\delta_{m})+
    \frac{\dot{\delta}_{m}^{2}}{1+\delta_{m}}, \label{deltadubbledot}
\end{eqnarray} where the dot denotes the derivative with respect to time. To estimate the second term of the above equation, ($\dot{H}-\dot{h}$),
we use Eqs. (\ref{Fried1}), and (\ref{Fried2}) for the background
and local regions, i.e.,
\begin{eqnarray}
&&  \dfrac{\ddot{a}}{a}= \dfrac{\Lambda}{3}-\dfrac{\rho_{m}}{6}-\dfrac{\eta}{3}\rho_{m}^2.
    \label{adubbledot}\\
&&\dfrac{\ddot{a}_{p}}{a_{p}}=\dfrac{\Lambda}{3}-\dfrac{\rho_{m}^c}{6}-\dfrac{\eta}{3}(\rho_{m}^c)^{2}. \label{apduubeldot-delta}
\end{eqnarray}
Despite that one may expect $\eta$ differs inside and outside of the spherical region, for simplicity here we suppose that $\eta^{c}=\eta$.
From Eqs. (\ref{delta}), (\ref{adubbledot}), and (\ref{apduubeldot-delta}), one comes to
\begin{eqnarray}
    \dot{H}-\dot{h}&=&\dfrac{1}{6}\rho_{m}\delta_{m} +\dfrac{2}{3}\eta \rho_{m}^2 \delta_{m}-H^{2}+h^{2}.
    \label{Hdot-hdot}
\end{eqnarray}
Now, by combining Eqs. (\ref{Hdot-hdot}) in (\ref{deltadubbledot}), and using (\ref{deltadot}), one get the following differential equation for the density contrast modified by EMSG model parameter $\eta$
\begin{eqnarray}
    &&\ddot{\delta}_{m}+2H\dot{\delta}_{m}-\dfrac{1}{2}\rho_{m}\delta_{m}-2 \eta\rho_{m}^{2}\delta_{m}=0.
    \label{deltaduubeldot2}
\end{eqnarray}
The differential equation above is written in terms of a time derivative, and as usual, we should re-write it in the form of the derivative in terms of the scale factor $a$. To do so, we introduce the following equations in which the prime represents the derivative in terms of the scale factor $a$
\begin{equation}
    \dot{\delta}_{m}=\delta_{m}^{\prime}aH, \quad \ddot{\delta}_{m}
    =\delta_{m}^{\prime\prime}a^{2}H^{2}+aH^{2}\delta^{\prime}_{m}+a \dot{H} \delta^{\prime}_{m},
    \label{prim}
\end{equation}
 Using Eqs. (\ref{Fried1}) and (\ref{adubbledot})
\begin{eqnarray}
\dot{H}&=&-\dfrac{1}{2}\rho_{m}-\dfrac{1}{2}\eta \rho_{m}^{2}.
\label{Hdot}
\end{eqnarray}
and substituting it in Eq. (\ref{prim}), we arrive at
\begin{equation}
\ddot{\delta}_{m}
=\delta_{m}^{\prime\prime}a^{2}H^{2}+aH^{2}\delta^{\prime}_{m}-\dfrac{a}{2}\rho_{m}\left(1+\eta \rho_{m}\right)\delta^{\prime}_{m}.
\label{prim2}
\end{equation}
Now by combining Eq.(\ref{deltaduubeldot2}) with Eqs. (\ref{prim}), and (\ref{prim2}), after ignoring the terms containing $O(\delta_{m}^2)$, and $ O({\delta^{\prime}_m}^2)$, we have
\begin{eqnarray}\label{deltafora}
&&\delta^{\prime\prime}_{m}+\dfrac{3}{a}\delta^{\prime}_{m}-\dfrac{1}{2aH^2}\rho_m \left( 1+\eta \rho_m\right)\delta^{\prime}_{m}\nonumber\\&&
- \dfrac{1}{2a^2H^2}\rho_m \left( 1+4\eta \rho_m\right)\delta_{m}=0.
\end{eqnarray}
Finally by putting $H^{2}$ from Eq. (\ref{Fried1}) in Eq. (\ref{deltafora}), we deal with the following differential equation
\begin{eqnarray}
&&\delta^{\prime\prime}_{m}+\dfrac{3}{a}\delta^{\prime}_{m}-\dfrac{3}{2a}\times\dfrac{ \rho_m+\eta \rho_m^2}{\Lambda+\rho_m+\dfrac{\eta}{2}\rho_m^2}\delta^{\prime}_{m}\nonumber\\&&
-\dfrac{3}{2a^2}\times\dfrac{ \rho_m+4\eta \rho_m^2}{\Lambda+\rho_m+\dfrac{\eta}{2}\rho_m^2}\delta_{m}=0.
\nonumber\\
\label{betadelta}
\end{eqnarray}
One openly can see that in the absence of the EMSG model parameter ($\eta=0$), and cosmological constant ($\Lambda=0$), the perturbed equation for the density contrast, $\delta_m$, in the linear regime, coincides with the ones in the GTR cosmology \cite{Abramo}
\begin{eqnarray}\label{deltagr}
&&\delta^{\prime\prime}_{m}+\dfrac{3}{2a}\delta^{\prime}_{m}-\frac{3\delta_{m}}{2a^{2}}=0,
\end{eqnarray}
In Fig. \ref{Fig3}, within the range $10 < z < 20$ of redshift,
we have plotted the matter density contrasts as a function
of $z$ for the different values of the EMSG model parameter $\eta$.
We observe that the EMSG cosmology leaves distinguishable imprints from $\Lambda$CDM on the density contrast of matter.
More exactly, in the presence of $\eta$, the density contrast of matter starts growing from its initial value so that in an expanding universe, its growth is faster than the $\Lambda$CDM profile. 
However, Fig. \ref{Fig3} reflects an inconsistency in the behavior of $\eta<0$. Although we see for $\eta > 0$ as it decreases, the growth of matter disturbances becomes slower and eventually goes to the $\Lambda$CDM model, for $\eta<0$, the opposite happens, and the matter perturbations grow faster.
In the light of Fig. \ref{Fig3}, we expect that with the increase of the $\eta$ parameter, the matter perturbations will grow faster so that the behavior of $\eta<0$ is nothing but an anomaly \footnote{Already also discussed in Ref. \cite{Akarsu} on the unsatisfactory behavior of $\eta<0$ in other contexts and the possibility of its discarding in case of repeating such an unrealistic physical result.}.
Then, from now on, we ignore it in our considerations and limit ourselves to the positive range of $\eta$. In general, quadratic contribution added to gravity from the side of the matter in the framework of EMSG is the principal reason for amplifying the growth of matter perturbations relative to the $\Lambda$CDM model.

\begin{figure}[t]
\epsfxsize=7.5cm \centerline{\epsffile{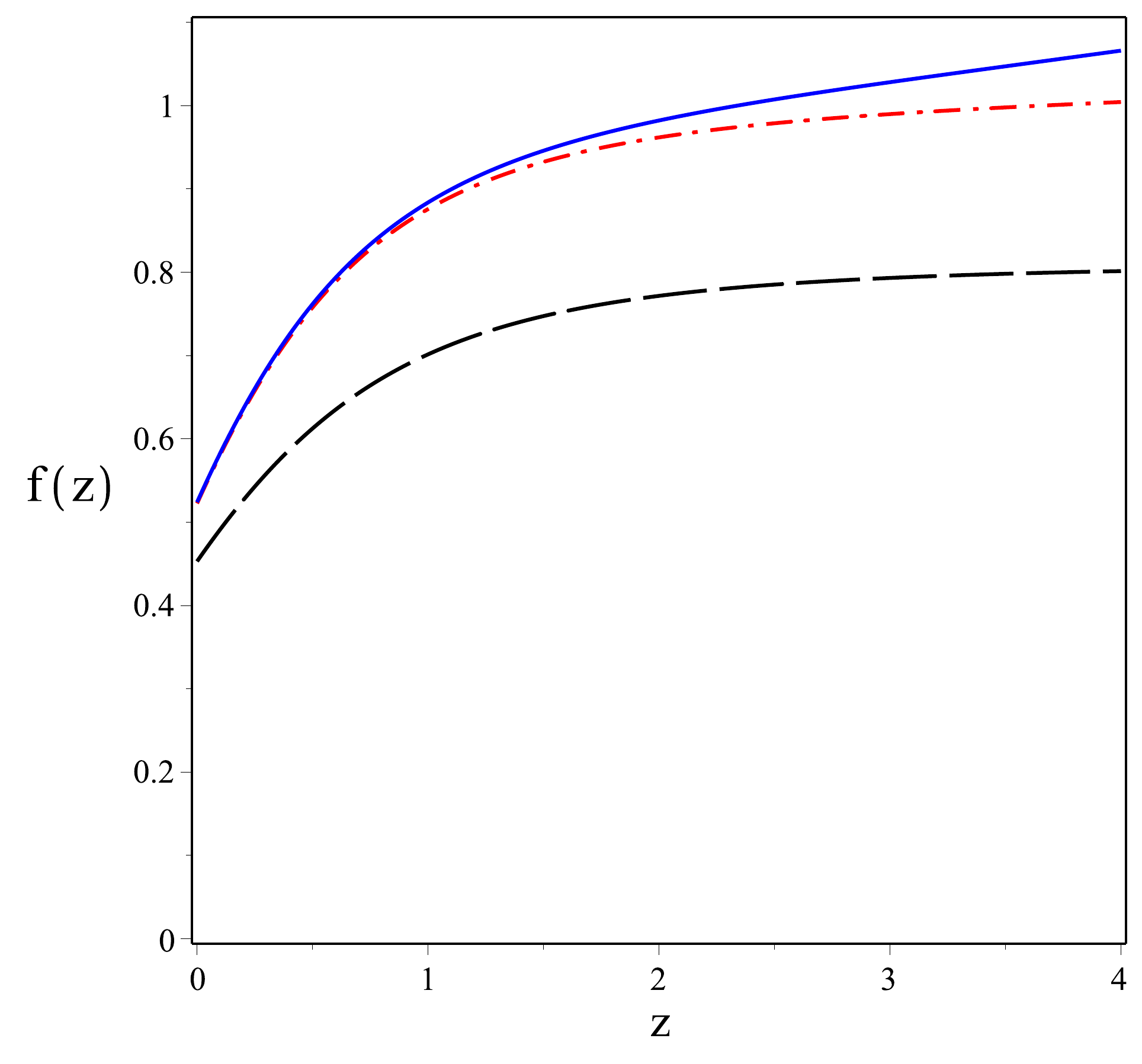}}
\caption{The evolution of the growth function for different values of the EMSG model parameter $\eta$: $\eta=0$ ($\Lambda$CDM, black-long-dashed), $\eta= 10^{-9}$ (red-dashed-dotted), and $\eta=10^{-8}$ (blue-solid).}
\label{fig4}
\end{figure}

In this regards, by serving the growth function as \cite{Peebles}
\begin{equation}\label{growthrate}
f(a)=\dfrac{dlnD}{dlna},\      \     \     \    \
D(a)=\dfrac{\delta_{m}(a)}{\delta_{m}(a=1)},
\end{equation}
one can investigate the growth rate of matter perturbations.
Let us recall that in the absence of the EMSG model parameter
($\eta= 0$), the growth function is a constant of unity. In Fig. \ref{fig4}, we display the growth function in terms of the redshift parameter. 
First, like the $\Lambda$CDM model, we observe that the amplitude of perturbations in high redshifts approaches the unity, while it starts to decrease at low redshifts.
It means that the role of the EMSG model parameter is not in conflict with $\Lambda$ to reduce the growth function from unity. Besides, the current value of $f(z)$ crucially depends on the $\eta$ parameter so that by increasing it, $f(z)$ deviates more from $\Lambda$CDM model.

\begin{figure}[t]
\epsfxsize=7.5cm \centerline{\epsffile{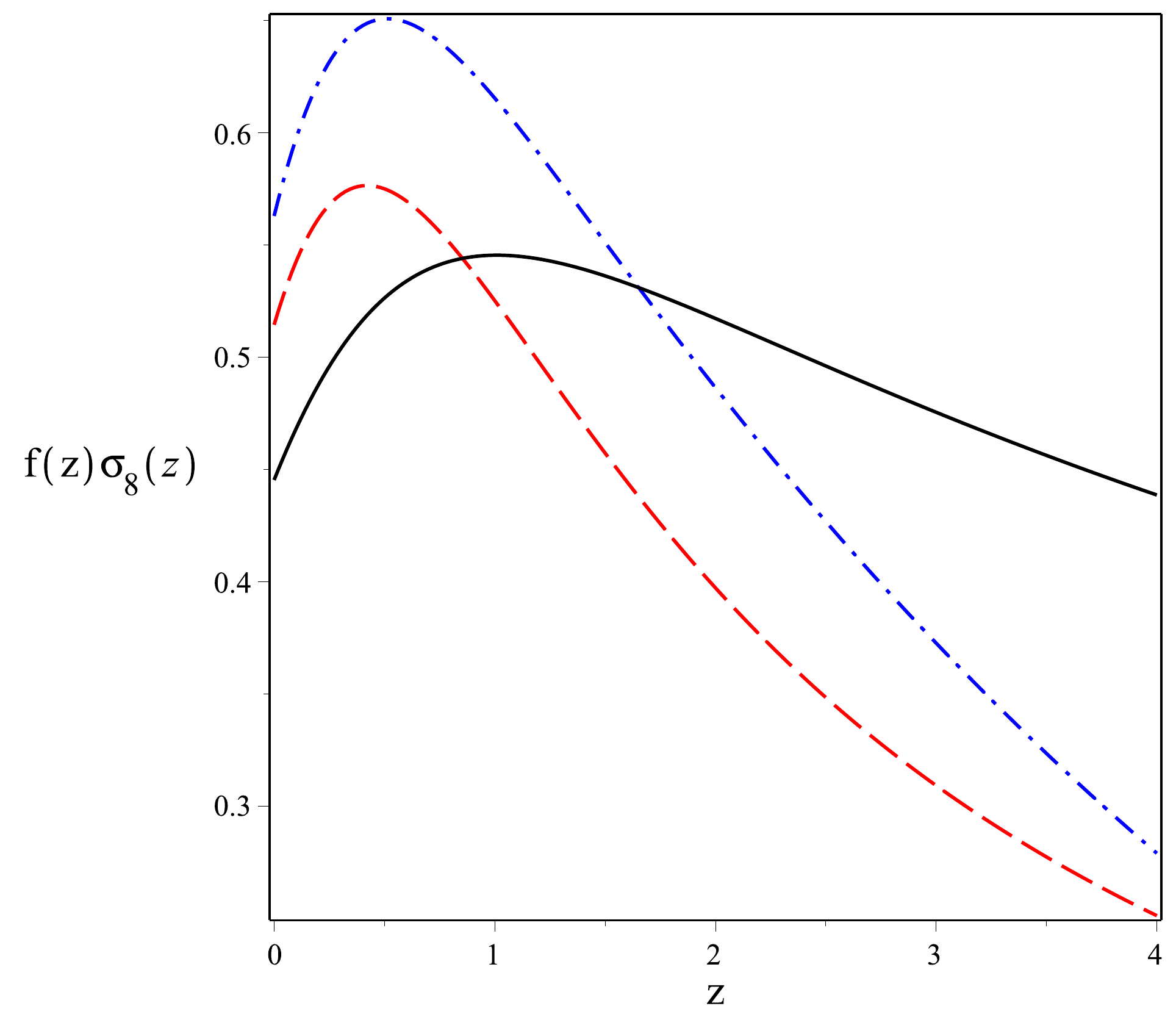}}
\caption{The behavior of $f(z)\sigma_{8}(z)$ for different values of the EMSG model parameter $\eta$: $\eta=0$ ($\Lambda$CDM, black-solid), $\eta=10^{-8}$ (red-dashed), and $\eta= 10^{-6}$ (blue-dashed-dotted). } \label{fig5}
\end{figure}

There is also another possibility of measuring the growth rate matter perturbations, which come from the redshift-space distortion (RSD) of the clustering pattern of galaxies.
This distortion is caused by the odd velocity of the inward collapse motion of the large-scale structure, which is directly linked to the growth rate of the matter density contrast $\delta_{m}$ \cite{Kaiser}. The recent surveys of galaxy redshift,  have
provided bounds on the growth rate $f(z)$ or $f(z)\sigma_{8}(z)$
in terms of the redshift where $f(z)$ comes from Eq. (\ref{growthrate}), and $\sigma_{8}$ is the rms (root mean square) amplitude of $\delta_{m}$ at the comoving scale $8h^{-1}Mpc$ \cite{Tsujikawa, Nesseris} and can be written as \cite{Nesseris2}
\begin{equation}
    \sigma_{8}(z)=\dfrac{\delta(z)}{\delta(z=0)}\sigma_{8}(z=0).
    \label{sigma8z}
\end{equation}
With this suppose that $\sigma_{8}(z=0)=0.983$ \footnote{Throughout this manuscript the values of Hubble constant $H_0$ and the $\sigma_8$
    are fixed at the Planck-$\Lambda$CDM measurements as
    $H_0 = 67.66 \pm 0.42 km s^{-1} Mpc^{-1}$, and $\sigma_8 = 0.983\pm 0.0060$ (CMB power spectra+CMB lensing+BAO) \cite{Nesseris}.} \cite{Nesseris},
we release the redshift evolution of $f(z)\sigma_{8}(z)$ for
different values of $\eta$ in Fig. \ref{fig5}.
While, in small redshifts ($z<1$) the EMSG model with a larger $\eta$ predicts a larger value of the cosmological growth rate, by going to large redshifts the behavior of models turns the same.
Indeed by adding the quadratic contributions to gravity from matter terms in EMSG model, we observe from Fig. \ref{fig5} that in small redshifts, the growth rate function is larger than $\Lambda$CDM model. With the presence of EMSG model parameter $\eta$, we also see that $f \sigma_{8}$ reaches the maximum value at redshifts smaller than $\Lambda$CDM. Namely, the large structures form in the EMSG model later than the standard counterpart.

\section{Number of galaxy cluster in EMSG cosmology}\label{MN}
In addition to the evolution of matter density contrast, another quantity that we are interested in surveying in the framework of EMSG cosmology is the number count of collapsed objects. This section indeed studies the distribution of the number density of collapsed objects of a given mass range in the framework of EMSG cosmology. The collapsed objects, which, in essence, are the chief source of large-scale structure formation of the universe, are called the DM halos. Besides, the baryonic matter, due to gravitational attraction, follows the DM distribution. In this way, tracing the distribution of DM haloes becomes possible by seeing the distribution of galaxy clusters. To do so, i.e., the investigation of the number distribution of the collapsed objects or the galaxy clusters along the redshift commonly is employed a semi-analytic approach known as the Press-Schechter formalism \cite{William}.
From the view of the mathematical formulations of the halo mass function, the matter density field usually should be enjoyed the Gaussian distribution.

The comoving number density of the gravitationally collapsed objects (equivalent to galaxy clusters) at a certain redshift $z$ having mass from $M$ to $M + dM$ is given by the following analytical formula  \cite{Liberato}
\begin{equation}
    \dfrac{dn(M,z)}{dM}=-\dfrac{\rho_{m,0}}{M}\dfrac{d\ln \sigma(M,z)}{dM} f(\sigma(M,z)),
    \label{massfunction}
\end{equation}
where $\rho_{m,0}$, $\sigma(M, z)$, and $f(\sigma)$ respectively denote the present matter mean density of the universe, the rms of density fluctuation in a sphere of radius $r$ surrounding a mass $M$, and the mathematical mass function proposed by Press and Schechter \cite{William}, as follow
\begin{equation}
f_{PS}(\sigma)=\sqrt{\dfrac{2}{\pi}} \dfrac{\delta_{c}(z)}{\sigma(M,z)}\exp\left[ -\dfrac{\delta_{c}^{2}(z)}{2\sigma^{2}(M,z)}\right].
\label{fsigma0}
\end{equation}
Subscript ''PS'', refers to Press and Schechter. Note that $\delta_{c}(z)$ in the mass function above is the critical density contrast above which structures collapse. By serving the linerised growth factor $D(z) = \delta_{m}(z)/\delta_{m}(z=0)$, as well as the rms of density fluctuation at a fixed length $r_{8} = 8h^{-1} Mpc$, then $\sigma(M, z)$, express as follow
\begin{equation}
    \sigma(z,M)= \sigma(0,M_{8})\left(\dfrac{M}{M_{8}} \right)^{-\dfrac{\gamma}{3}} D(z),
    \label{sigmazm}
\end{equation}
where the index $\gamma$ is reads as \cite{Mukherjee,Mukherjee2}
\begin{equation}
    \gamma=(0.3\Omega_{m,0}h+0.2)\left[ 2.92+\dfrac{1}{3} \log\left(  \dfrac{M}{M_{8}}\right) \right],
    \label{gamma}
\end{equation} and $M_{8}=6\times10^{14}\Omega_{M}^{(0)}h^{-1}M_{\bigodot}$ is the mass inside a sphere of radius $r_{8}$ ($M_{\bigodot}$ is the solar mass) \cite{Viana}.

\begin{figure}[t]
    \epsfxsize=7.5cm \centerline{\epsffile{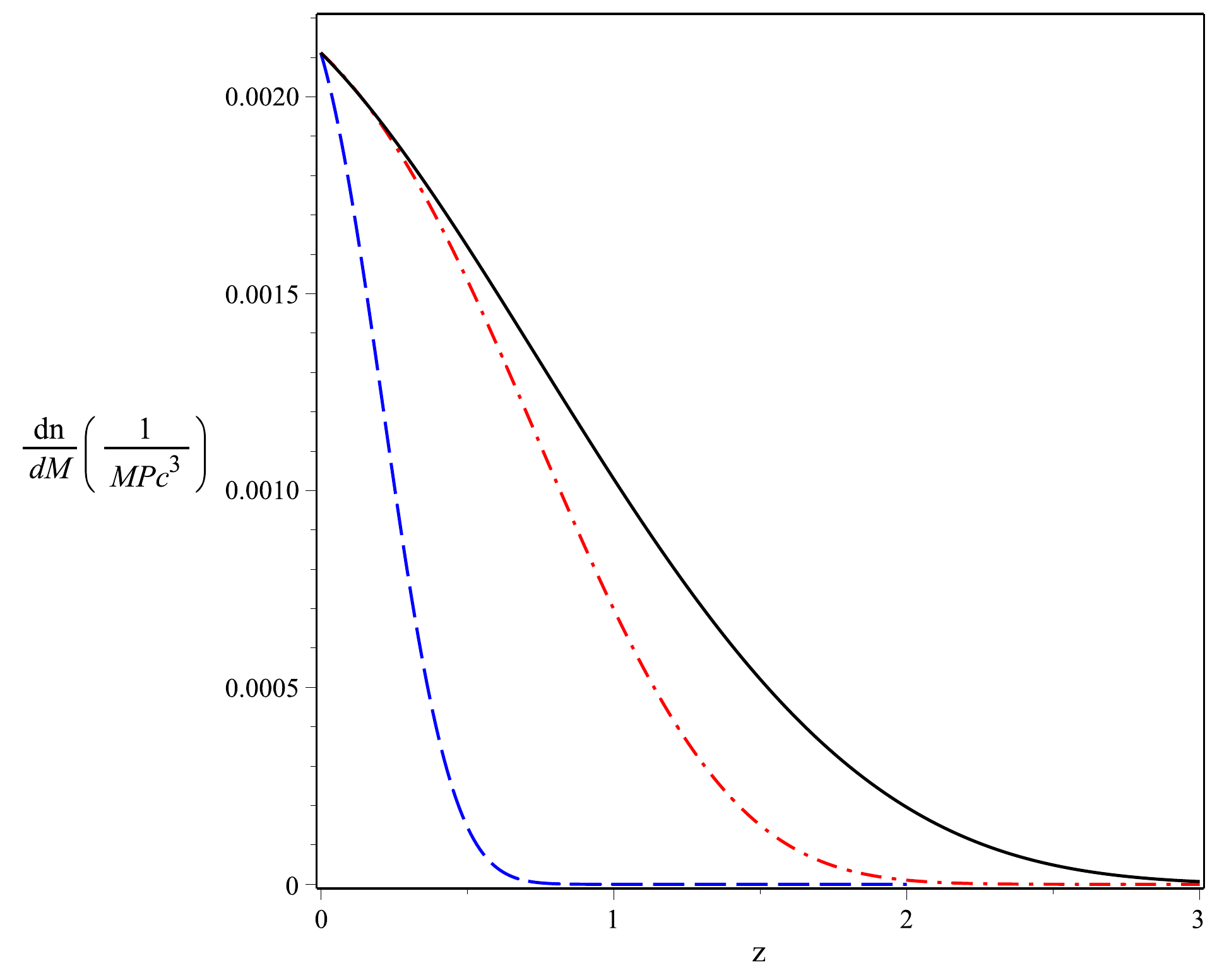}}
    \caption{The evolution of mass function for objects with mass $M=10^{13}(h^{-1}M_{\bigodot})$, and different values of the EMSG model parameter $\eta$: $\eta=0$ ($\Lambda$CDM, black-solid), $\eta=10^{-5}$ (red-dashed-dotted), and $\eta=10^{-3}$ (blue-dashed). In general, for objects with the mass range: $10^{13}-10^{16}h^{-1}M_{\bigodot}$, we are faced with this same qualitative behavior.}
    \label{fig6}
\end{figure}
\begin{figure}[t]
    \epsfxsize=7.5cm \centerline{\epsffile{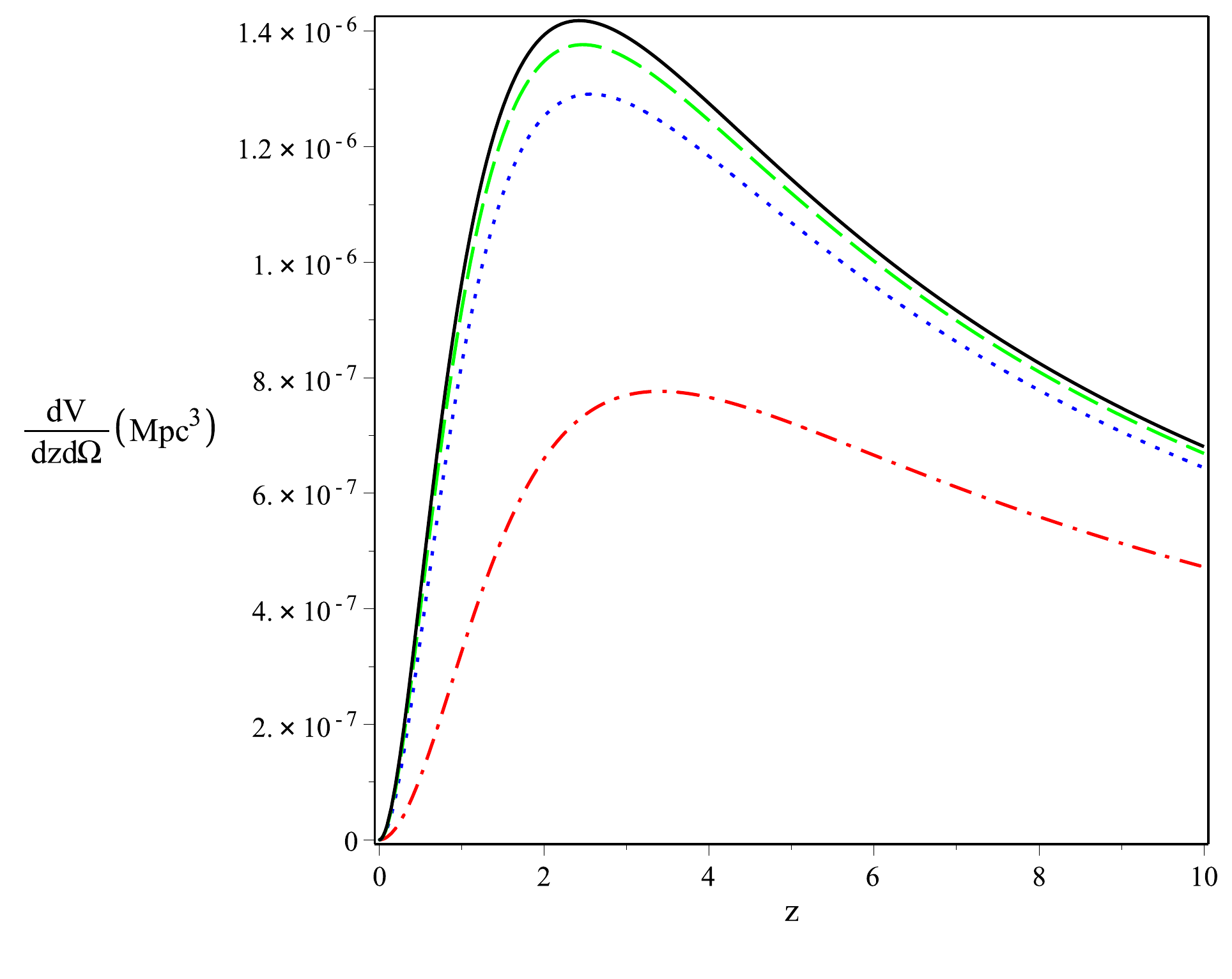}}
    \caption{The behavior of the comoving volume element with different values of the EMSG model parameter $\eta$: $\eta=0$ ($\Lambda$CDM, black-solid curve), $\eta=10^{-5}$ (green-dashed curve), $\eta=10^{-4}$ (blue-dotted curve), and $\eta=10^{-3}$ (red-dashed-dotted curve).} \label{fig7}
\end{figure}
\begin{figure}[t]
\epsfxsize=7.5cm \centerline{\epsffile{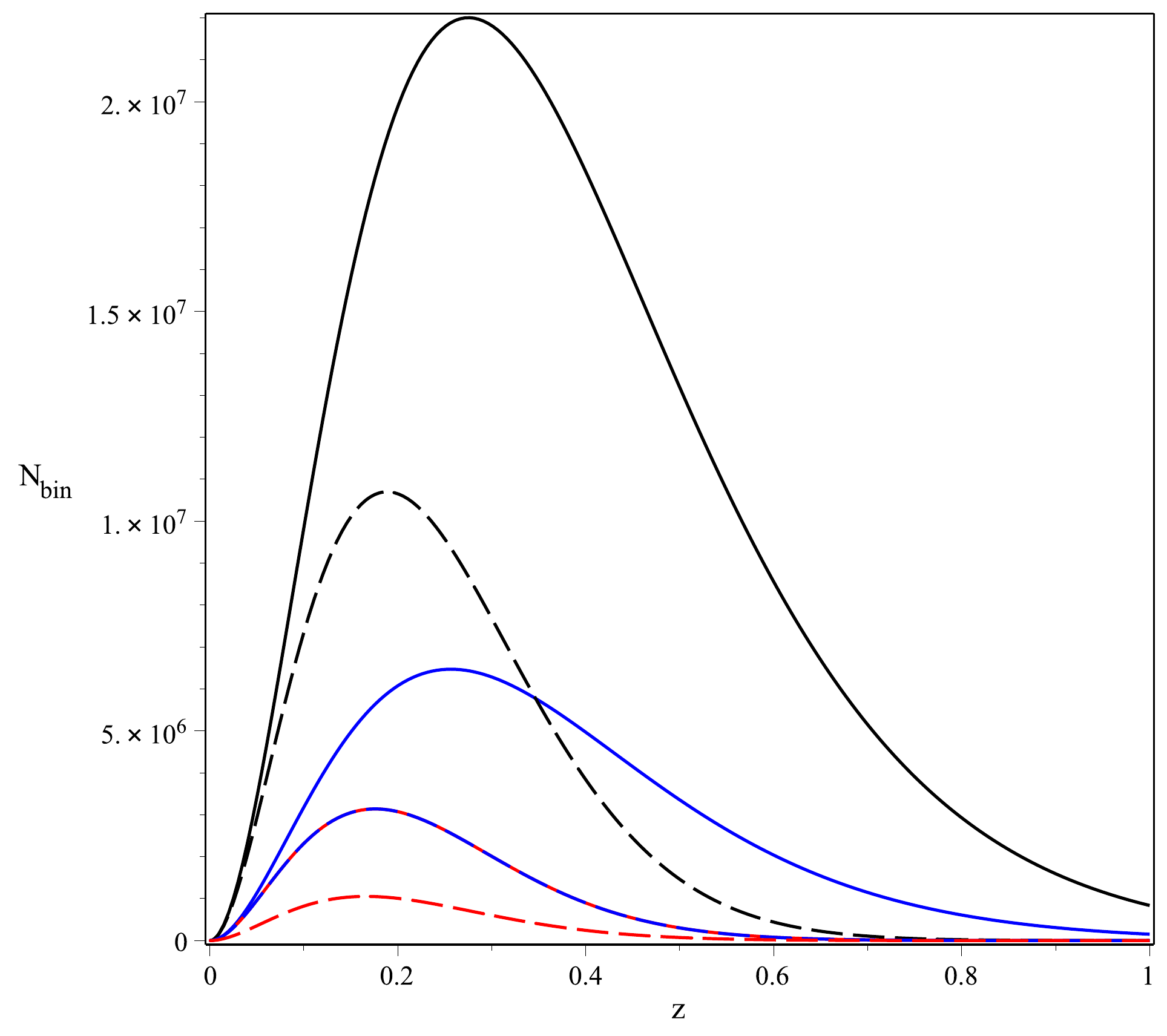}}
    \caption{The evolution of cluster number count with redshift from the $\Lambda$CDM (solid curves), and EMSG with $\eta=10^{-5}$ (dashed curves), for objects with mass within the range: $10^{12}<M/(h^{-1}M\bigodot)<10^{13}$ (black), $10^{13}<M/(h^{-1}M\bigodot)<10^{14}$ ( blue), $10^{14}<M/(h^{-1}M\bigodot)<10^{15}$ (red).} \label{fig8}
\end{figure}
We first in Fig. \ref{fig6}, display the redshift evolution of
mass function ($dn/dM (1/Mpc^3)$) of objects with mass $10^{13}
h^{-1}M_{\bigodot}$ for different values of EMSG model parameter
$\eta$.  It is important since is one of the quantities involved
in the number of DM halos. Fig. \ref{fig6} explicitly shows that
the presence of the $\eta$ causes the growth of the mass function
to start around redshift less than the $\Lambda$CDM model, i.e.,
the halo abundance is formed later. Namely, as the parameter
$\eta$ becomes larger, the halo abundance grows rapidly at lower
redshifts.

Finally, to survey the number of DM halos in EMSG cosmology, we
employ the effective number of collapsed objects between a given
range of mass bin $M_{inf} < M < M_{sup}$ per unit of redshift
\begin{equation}
\mathcal{N}^{bin}\equiv\dfrac{dN}{dz}=\int_{4\pi} d\Omega \int^{M_{sup}}_{M_{inf}}\dfrac{dn}{dM}\dfrac{dV}{dzd\Omega}dM,
\label{numbercounts}
\end{equation}
where $\dfrac{dV}{dzd\Omega}$ is the comoving volume element and is defined as
\begin{equation}
\dfrac{dV(z)}{dz d\Omega}=\dfrac{r^{2}(z)}{H(z)},~~~r(z)=\int^{z}_{0} H^{-1}(x)dx
\label{volume}
\end{equation}
Here $r(z)$ denotes the comoving distance. We depict the redshift
evolution of the comoving volume element ($dV/dz d\Omega(Mpc^3)$)
with various values of the EMSG model parameter $\eta$ in Fig.
\ref{fig7}. As we see the comoving volume element becomes smaller
in the presence of $\eta$ parameter. It is worth noting that the
comoving volume element just depends on the cosmological
background and the growth factor of the perturbation $D(z)$ does
not affect it. By setting a certain value for EMSG model parameter
$\eta$, as well as taking $\Lambda$CDM into of account, for
various mass bins $[M_{inf} ,M_{sup}]$ from $10^{12}
h^{-1}M_\odot$ to  $10^{15} h^{-1}M_\odot$, we display in Fig.
\ref{fig8} the behavior of $\mathcal{N}^{bin}-z$ (the number count
in mass bins). It has two clear messages. First of all, the
cluster number count in the presence of the EMSG model parameter
is less than its standard counterpart. Second, the more massive
structures are less abundant and form at later times which is in
agreement with what we expect from the hierarchical model of
structure formation.

\section{Conclusion}\label{CD}
It is well-known that utilizing the SC formalism of the matter's overdensity is a suitable method to study the effects of ``Modified Gravity'' on the large-scale structure of the universe. 
In the present work, by serving SC formalism, we have studied the linear evolution of matter's overdensity in the cosmology framework arising from EMSG, which affects the behavior of standard $\Lambda$CDM in the early universe due to inducing quadratic contributions from the matter to gravity. 
These corrections affect the expansion history of
the universe, too. In practical phenomenological terms, one may
imagine it as a potential extension of the $\Lambda$CDM model of
cosmology. More precisely, using Friedmann dynamics equations (in
the presence of a cosmological constant) modified with terms
containing EMSG model parameter $\eta$, and by taking into account
these new effects on the growth of perturbations, we have
investigated the structure formation beyond $\Lambda$CDM.

It is well-known from standard cosmology that the universe in the past around a redshift $z_{tr}$, has been switched  from decelerated
phase to an accelerated phase, and it continues today.
We found that the presence of $\eta$ terms in dynamics of EMSG cosmology affects the value of $z_{tr}$ so that it increases (decreases)
with $\eta>0$ ($\eta<0$) meaning that our universe experiences
this phase transition sooner (later) than $\Lambda$CDM.
Our evaluation of the evolution of the matter density contrast for different values of $\eta$ ruled out case $\eta<0$ and showed that the matter density contrast grows up faster than the $\Lambda$CDM profile. It is, in essence, due to nonlinear matter extensions appearing in the dynamics of EMSG cosmology. The growth function increases in the presence of the EMSG model parameter $\eta$, too.
In this regard, we have investigated $f \sigma_{8}$ for EMSG
cosmology and found that in the presence of $\eta$ parameter, $f
\sigma_{8}$ reaches the maximum value at smaller redshifts. Namely, the large-scale structures in the framework of EMSG cosmology form later compared to $\Lambda$CDM.

The modifications caused by EMSG cosmology affect the number of DM
halos so that the presence of $\eta$ makes halo abundances smaller
than $\Lambda$CDM. In other words, for larger $\eta$, the mass
function starts to grow in smaller redshifts meaning that the halo
abundance is formed later relative to $\Lambda$CDM. The comoving
volume element becomes smaller in EMSG cosmology, too. 
In addition, in light of analyzing the number counts in mass bins 
for the cosmology model at hand, two results were obtained. First,
the number of galaxy clusters in EMSG is less than $\Lambda$CDM
model. Second, we found that in EMSG, more massive structures are
less abundant and form at later times, as expected from 
the hierarchical model of structure formation.

The present study shows that the evolution of linear perturbation
reacts to the EMSG model parameter and it leaves distinguishable
imprints from $\Lambda$CDM. For this reason, the outputs of such
studies would be helpful to constrain the models based on future
observations like type Ia supernovas and baryon acoustic
oscillations, etc.

\end{document}